\documentclass[12pt,a4paper]{article}
\usepackage{epsfig}
\newcommand{\beq}{\begin{equation}}
\newcommand{\eeq}{\end{equation}}
\newcommand{\beqn}{\begin{eqnarray}}
\newcommand{\eeqn}{\end{eqnarray}}
\newcommand{\beqns}{\begin{eqnarray*}}
\newcommand{\eeqns}{\end{eqnarray*}}

\def\PL{{\it Phys. Lett.}}
\def\PR{{\it Phys. Rev.}}

\def\babar{\mbox{\slshape B\kern-0.lem{\smaller A}\kern-0.lem 
    B\kern-0.lem{\smaller A\kern-0.2em R}}\xspace}

\begin{document}
\normalsize
\begin{flushright}
\today \\
\end{flushright}

\begin{center}
\vspace{2.5cm}
{\Large
\bf

Two-photon exchange model for production of neutral vector meson pairs
in $e^+e^-$ annihilation}

\vspace{1.cm}

\begin{large}
M.~Davier$^{\,\mathrm a}$,
M.~Peskin$^{\,\mathrm b}$,
and A.~Snyder$^{\,\mathrm b}$\footnote
{
	E-mail: 
	davier@lal.in2p3.fr,
	mpeskin@slac.stanford.edu,
	snyder@slac.stanford.edu
} \\
\end{large}
\vspace{0.5cm}
{\small \em $^{\mathrm a}$Laboratoire de l'Acc\'el\'erateur Lin\'eaire,\\
IN2P3-CNRS et Universit\'e de Paris-Sud 11, 91898 Orsay, France}\\
\vspace{0.1cm}
{\small \em $^{\mathrm b}$Stanford Linear Accelerator Center, Stanford,
			  California 94309, USA}\\

\end{center}

\vspace{1.5cm}

\begin{abstract}
A vector-dominance two-photon exchange model is proposed to explain the recently
observed production of $\rho^0\rho^0$ and $\rho^0\phi$ pairs in $e^+e^-$
annihilation at 10.58 GeV with the BaBar detector. All the observed features
of the data ---angular and decay distributions, rates--- are in agreement
with the model. Predictions are made for yet-unobserved final states.

\end{abstract}

\section{Introduction}

So far all observed hadronic processes in $e^+e^-$ annihilation are in 
agreement with one-photon exchange leading to $C=-1$ final states. However,
the BaBar Collaboration has recently presented the first measurement of 
$e^+e^-$ annihilation into pairs of neutral vector mesons at 10.58 GeV 
centre-of-mass energy~\cite{babar}. The final states $\rho^0~\rho^0$ and 
$\rho^0~\phi$ are observed with cross sections
\beqn
 \sigma(e^+e^-\rightarrow \rho~\rho) &=&
  (20.7 \pm 0.7_{\rm stat} \pm 2.7_{\rm syst})~{\rm fb}, \\
 \sigma(e^+e^-\rightarrow \rho~\phi) &=&
  (5.7 \pm 0.5_{\rm stat} \pm 0.8_{\rm syst})~{\rm fb}
\eeqn
in the centre-of-mass angular range $|\cos{\theta}|<0.8$. These processes
characterized by $C=+1$ final states cannot originate from one-photon
exchange followed by quark-antiquark fragmentation.

An explanation for these processes is that they arise from $e^+e^-$
annihilation to two virtual photons, with each virtual photon converting
to a vector meson.  The rates predicted by this mechanism can be computed
simply and unambiguously using the effective vector meson-photon
couplings determined from the meson leptonic widths.  In this note, we
present the results for the cross sections and show that they are of
the correct size to explain the cross section values obtained by BaBar.

\section{The model}

We consider the generic process $e^+e^- \rightarrow V_1V_2$, where $V_1$ and 
$V_2$ are neutral $C=-1$ vector mesons. It is assumed to proceed  with an
intermediate $e^+e^- \rightarrow \gamma~\gamma$ process, the two photons 
converting into $V_1$ and $V_2$, with effective couplings $e/f_1$ and 
$e/f_2$. 

The cross section in the narrow-width approximation 
can be simply written in terms of the Mandelstam 
variables $s=(k+k')^2$, $t=(k-p_1)^2$, and $u=(k-p_2)^2$, where $k$, 
$k'$, $p_1$, and $p_2$ are the incoming electron, positron, and outgoing 
vector meson 4-momenta, respectively. In the $e^+e^-$ centre-of-mass, one gets

\beq
\label{xsection}
    \frac {d\sigma_{V_1V_2}}{d\cos{\theta}} = 
    \left (\frac {e}{f_1} \right )^2 \left (\frac {e}{f_2} \right )^2 
  \frac {d\sigma_{\gamma_1^* \gamma_2^*}}{d\cos{\theta}}~(m_{V_1}^2,m_{V_2}^2)
\eeq
with
\beq
\label{gamma*}
\frac {d\sigma_{\gamma_1^* \gamma_2^*}}{d\cos{\theta}}~(m_{V_1}^2,m_{V_2}^2) = 
    \frac {\pi\alpha^2}{s} \frac {2|\overrightarrow{p}|}{\sqrt{s}}
   ~\frac {2(m_{V_1}^2+m_{V_2}^2)sut+(t^2+u^2)(ut-m_{V_1}^2m_{V_2}^2)}{u^2t^2},
\eeq
and where $\theta$ is the angle between the electron and the $V_1$ momenta, 
and $\overrightarrow{p}$ the meson momentum. The last factor reduces 
to the familiar 
$t/u+u/t~=~2(1+\cos^2{\theta})/\sin^2{\theta}$ when $V_1$ and $V_2$ 
are made massless. For $m_{V_1}=m_{V_2}=M_Z$, Eq.~(\ref{xsection}) is 
consistent with the well-known expression for the vector part of the
$e^+e^- \rightarrow Z~Z$ cross section~\cite{brown}.

For masses which are small compared to $\sqrt{s}$, the angular distribution 
differs little from the massless case. To leading order in $m_V^2/s$, 
for the $VV$ case, one obtains:

\beq
\label{angular}
 \frac {d\sigma_{VV}}{d\cos{\theta}} \sim \frac {a+\cos^2{\theta}}
       {b-\cos^2{\theta}},
\eeq
with
\beqn
  a &=& 1+\frac {8m_V^2}{s}, \\
  b &=& 1+\frac {4m_V^4}{s^2}.
\eeqn
For $m_V=m_\rho$ and $s=(10.58)^2$~GeV$^2$, $a=1.043$ and $b=1.00012$.

For the $VV$ final states involving 2 identical particles, the
angular distribution must be integrated over only one hemisphere,
while for $V_1V_2$ ($V_1 \neq V_2$), the integration should be done over
the full angular distribution. The measurements with BaBar are made from
in the $\cos{\theta}$ range from $-0.8$ to $0.8$.

To take into account the fact that the $\rho$ meson cannot be properly 
described in the narrow-width approximation and also to generalize to
any vector hadronic final state, it is advantageous to rewrite
Eq.~(\ref{xsection}) in a more general form:

\beq
\label{xsection_general}
    \frac {d\sigma_{V_1V_2}}{d\cos{\theta}} = 
    \left (\frac {\alpha}{3\pi} \right )^2 
    \int \frac {dm_1^2}{m_1^2} \int \frac {dm_2^2}{m_2^2}~
    R(m_1^2)~R(m_2^2)~
   ~\frac {d\sigma_{\gamma_1^* \gamma_2^*}}{d\cos{\theta}}~(m_1^2,m_2^2),
\eeq
where $R(m^2)=\sigma (e^+e^- \rightarrow {\rm hadrons})/\sigma_{pt}$.
Eq.~(\ref{xsection_general}) is a completely general
expression for the 2-virtual-photon cross section, valid as long as the
interference between the products of the 2 photons can be neglected.
All one has to do is plug in the contribution to R for any final state
and integrate. The narrow-width limit is obtained by taking
$R(m^2)= 9\pi/\alpha^2 ~\Gamma_{ee}^V~m_V~\delta(m^2-m_V^2)$.

\section{Results}

For the narrow vector mesons the leptonic widths can be directly used
to derive the $e/f_V$ values
\beq
 \Gamma_{ee}^V = \frac {\alpha}{3} \left (\frac {e}{f_V}\right )^2 m_V,
\eeq
while for the $\rho$ meson we integrate over the mass distribution, taking 
as input for $R(m^2)$ the fit to the annihilation data~\cite{cmd2,snd06}.

\begin{figure}[h] \centering
  \includegraphics[width=12.cm]{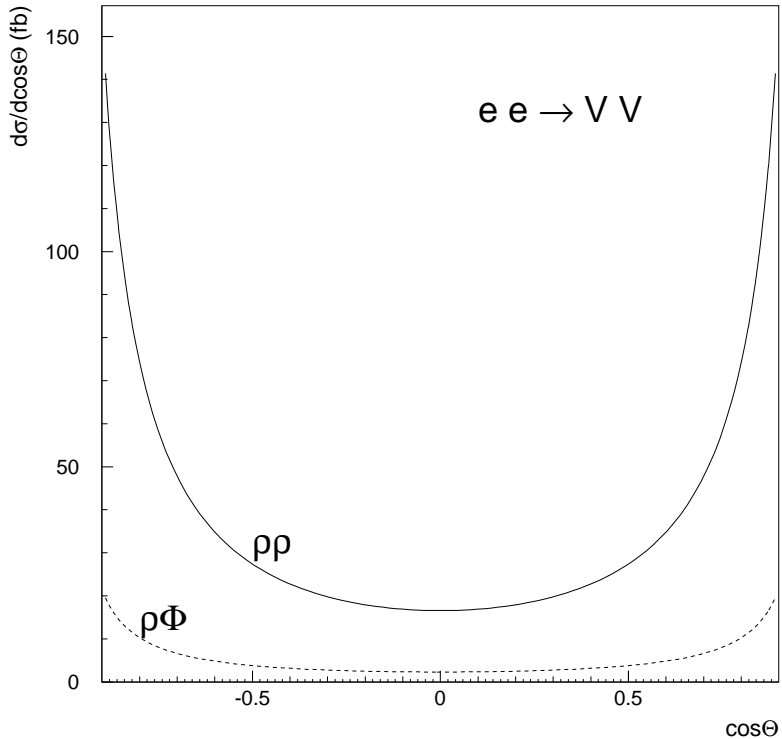}
  \caption{\label{angle} 
   The angular distribution for the processes $e^+e^- \rightarrow \rho~\rho$
   and $e^+e^- \rightarrow \rho~\phi$ at 10.58 GeV as predicted by the
   vector dominance model in the narrow width approximation.}
  \end{figure}

The angular distribution is given in Fig.~\ref{angle}. It is strongly
peaked along the beams, like the $e^+e^- \rightarrow \gamma \gamma$ cross
section. The behaviour predicted by 
Eqs.~(\ref{xsection},\ref{gamma*},\ref{angular})
is in good agreement with the BaBar data~\cite{babar}.

The model also allows one to compute the cross sections for mesons with
helicity $\lambda =\pm 1$ and $\lambda=0$. As expected the former is dominant,
{\it i.e.} the vector mesons are photon-like with transverse polarisation, 
leading to a $\sin^2{\theta_{\pi,K}}$ angular distribution in the 
$\rho \rightarrow \pi^+ \pi^-$ and $\phi \rightarrow K^+ K^-$ decays. 
This behaviour is also clearly observed in the BaBar data~\cite{babar}.

Integrating numerically Eq.~\ref{xsection_general} over $\cos{\theta}$ from
$-0.8$ to $+0.8$ and over the $\rho$ mass distribution, one obtains

\beqn
 \sigma_{\rho~\rho}^{\rm th} &=& (21.4 \pm 0.7)~{\rm fb}, \\
 \sigma_{\rho~\phi}^{\rm th} &=& (6.15 \pm 0.22)~{\rm fb}.
\eeqn

The BaBar measurements are given with a cut on the $\rho$ lineshape, 
retaining only events with a mass between 0.5 and 1.1 GeV. The $\phi$
window is likewise defined between 1.008 and 1.035 GeV.
The cross sections in the BaBar conditions are:

\beqn
 \sigma_{\rho~\rho}^{{\rm th},~{\rm ~mass ~cuts}} &=& (18.7 \pm 0.6)~{\rm fb}, \\
 \sigma_{\rho~\phi}^{{\rm th},~{\rm ~mass ~cuts}} &=& (5.3 \pm 0.2)~{\rm fb}.
\eeqn

The agreement with the BaBar results is good, thus confirming the
vector-dominance two-photon process as the dominant dynamics for $V_1^0V_2^0$
final states.

The calculation can be straightforwardly applied to other final states,
the corresponding predictions being summarized in Table~\ref{results}.
We emphasize that any interference between the hadronic products originating
from the two virtual photons has been neglected ---an approximation which
is expected to be less valid as the hadronic masses increases on the scale
of $\sqrt{s}$.

The value obtained for $e^+e^- \rightarrow J/\psi~J/\psi$ differs somewhat 
from the estimate given by Bodwin {\it et al}~\cite{braaten}. That 
calculation uses the 2-photon exchange model, with splitting of each 
photon in $c~ \overline{c}$.
The authors allow both pairings of the charm and anticharm quarks and
account for the interference between these diagrams using a non-relativistic
QCD analysis.  They find a cross section of $(6.6 \pm 3.0)$~fb, integrating the
full angular distribution, to be compared with $(2.38 \pm 0.15)$~fb for the
subset of diagrams that we consider.  While the new interference terms could
be important for $J/\psi$ production, since $2 m_{J/\psi}$ is close to
$\sqrt{s}$, they are not significant for the production of pairs of low-mass
states such as $\rho$ and $\phi$.

\begin{table}[t]
  \caption[.]{\label{results}
              Cross sections for the production of neutral vector meson pairs
              $V_1~V_2$ in $e^+e^-$ annihilation at 10.58 GeV. For the final 
              states containing $\rho$ mesons the results
              are given both with the narrow-width (NW) approximation and the
              full integration of the mass spectrum.
              The angular integration is restricted to the $\cos{\theta_{V_1}}$
              range from $-0.8$ to $+0.8$ for the results in 
              the last column. The values for the leptonic widths
              are taken from Refs.~\cite{cmd2,snd06,pdg}. The integral over
           the $\rho$ mass distribution according to Eq.~\ref{xsection_general}
              uses a fit to the annihilation data~\cite{cmd2,snd06}.}
  \begin{center}
\setlength{\tabcolsep}{0.0pc}
\begin{tabular*}{\textwidth}{@{\extracolsep{\fill}}lccccc} 
\hline\noalign{\smallskip}
 $V_1~V_2$ & $\Gamma_{V_2 \rightarrow ee}$ (keV) & $\sigma_{NW}$ (fb) & $\sigma$ (fb) & $\sigma _{|\cos \theta|<0.8}$ (fb) \\
\noalign{\smallskip}\hline\noalign{\smallskip}
 $\rho~\rho$         & $7.07\pm0.11$   &  $125.51$  & $115.69$ & $21.39$\\
 $\rho~\omega$       & $0.60\pm0.02$   &  $21.11$  & $20.27$ & $3.76$\\
 $\rho~\phi$         & $1.27\pm0.04$   &  $33.70$  & $32.39$ & $6.15$\\
 $\rho~J/\psi$       & $5.40\pm0.17$   &  $40.66$  & $39.61$ & $10.16$\\
 $\rho~\psi(2S)$     & $2.10\pm0.12$   &  $13.21$  & $12.89$ & $2.60$\\
 $\rho~\Upsilon(1S)$ & $1.31\pm0.03$   &  $7.83$  & $7.87$ & $5.45$\\
 $\phi~\phi$         & $-$             &  $2.23$  & $-$ & $0.44$\\
 $\omega~\phi$       & $-$             &  $2.83$  & $-$ & $0.54$\\
 $\omega~\omega$     & $-$             &  $0.89$  & $-$ & $0.16$\\
 $\phi~J/\psi$       & $-$             &  $5.06$  & $-$ & $1.46$\\
 $J/\psi~J/\psi$     & $-$             &  $2.38$  & $-$ & $1.13$\\
 $J/\psi~\psi(2S)$   & $-$             &  $1.49$  & $-$ & $0.78$\\
\noalign{\smallskip}\hline
  \end{tabular*}
  \end{center}
\end{table}

The predictions in Table~\ref{results} beyond the already measured 
$\rho~\rho$ and $\rho~\phi$ final states can be tested in the future as
some of them are within reach of the BaBar and Belle detectors with present
and foreseen luminosities.

\section{Contributions to $R$}

The computed cross sections can be extrapolated to lower energies in order to
estimate the contribution of these $C=+1$ two-photon processes in 
comparison to the dominant $C=-1$ one-photon annihilation into hadrons.
The latter one is the needed input to the calculations of hadronic vacuum 
polarization at lowest-order.

\begin{figure}[h] \centering
  \includegraphics[width=12.cm]{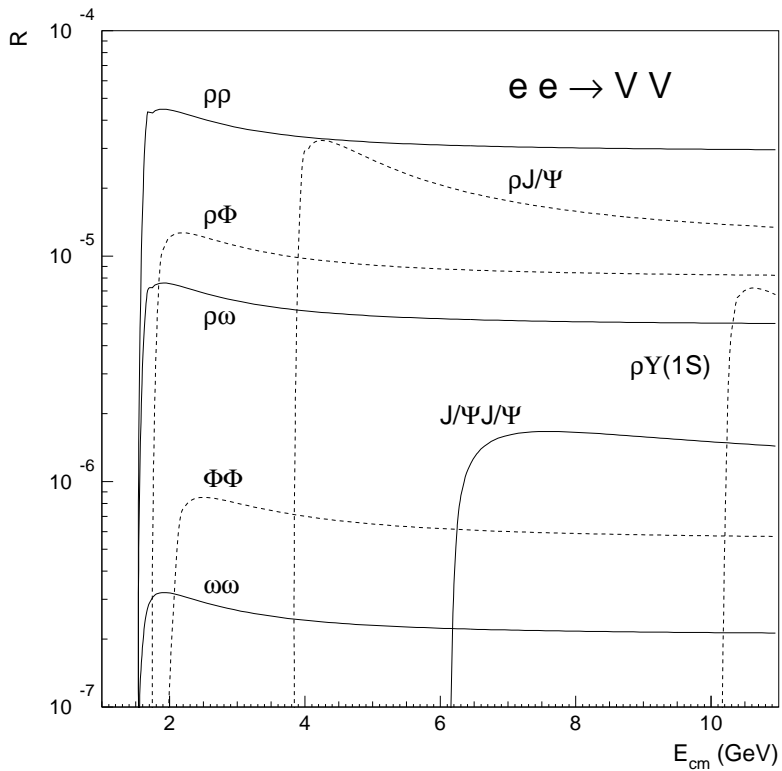}
  \caption{\label{R} 
   The contributions to $R$ from  the dominant processes 
   $e^+e^- \rightarrow V_1~V_2$ as predicted by the vector dominance model 
   in the narrow-width approximation and integrated in the 
   $[-0.8-0.8] ~\cos{\theta}$ range.}
  \end{figure}

Figure~\ref{R} shows the $\sqrt{s}$ dependence of the cross sections 
for the dominant processes, computed in the narrow-with approximation and
expressed relatively to the pointlike cross section ($R$). The angular 
integration is still performed from $\cos{\theta_{V_1}}=-0.8$ to $+0.8$, 
values which correspond to a typical experimental acceptance.
It is found that the contribution of these 2-photon exchange processes to
the measured $R$ is negligible at all energies. In particular,
their effect in hadronic vacuum polarization calculations (effect which is 
presently neglected), such as for the anomalous muon magnetic moment 
or the running of $\alpha$ is two orders of magnitude smaller than the
present experimental accuracy from the $R$ input values.

Interference effects between $C=-1$ and $C=+1$ amplitudes could occur 
at the $10^{-2}-10^{-3}$ level, but they cancel for charge-symmetric event 
detection. So their contribution is negligible.

\section{Conclusions}
The proposed two-photon exchange vector-dominance model for the processes 
$e^+e^- \rightarrow V_1~V_2$, where $V_1$ and $V_2$ are $C=-1$ neutral vector 
bosons, agrees with all the features of the BaBar data~\cite{babar} for 
the $\rho~\rho$ and $\rho~\phi$ final states and provides a good description 
of their rates.

The cross sections for other possible final states, $\rho~\omega$, 
$\rho ~J/\psi$, $\rho~\psi(2S)$, $\rho~\Upsilon(1S)$, $\phi~\phi$, 
$\omega~\phi$, $\omega~\omega$, $\phi~J/\psi$, $J/\psi~J/\psi$, and 
$J/\psi~\psi(2S)$ have been predicted. Most of them are within reach at the
B factories with foreseen luminosities.

The contamination of these two-photon exchange processes to the measured $R$ 
ratio is much smaller than any foreseeable experimental uncertainty on $R$. \\

Helpful discussions with S.~Brodsky and many BaBar colleagues are warmly
acknowledged.

\end{document}